\documentclass[useAMS,usenatbib]{mn2e}
\usepackage{graphicx}

\title[Ca\,{\sc ii k} interstellar observations towards early disc and halo 
stars - distances to IVCs and HVCs]{Ca\,{\sc ii k} interstellar observations 
towards early disc and halo stars - distances to intermediate and 
high-velocity clouds}

\author[J.V. Smoker et al.]
         {J.~V. Smoker$^{1,2}$\thanks{email: jsmoker@eso.org},
          B.~B. Lynn$^{2}$, W.~R.~J. Rolleston$^{2}$, H.~R.~M. Kay$^{3}$,
          E. Bajaja$^{4}$, \newauthor 
          W.~G.~L Poppel$^{4}$, F.~P. Keenan$^{2}$, 
          P.~M.~W Kalberla$^{5}$, C.~J. Mooney$^{2}$, \newauthor
          P.~L. Dufton$^{2}$, 
          R.~S.~I. Ryans$^{2}$
          \\
       $^{1}$European Southern Observatory,
            Alonso de Cordova 3107,
            Casilla 19001,
            Vitacura,
            Santiago 19, Chile
                 \\
       $^{2}$Astrophysics and Planetary Science Division,
            Department of Pure and Applied Physics,
            The Queen's University of Belfast, \\
            University Road, Belfast, BT7 1NN,
            U.K. \\
       $^{3}$Mullard Space Science Laboratory,
            University College London,
            Holmbury St Mary,
            Dorking,
            Surrey,
            RH5 6NT,
            U.K. \\
       $^{4}$Instituto Argentino de Radioastronomia,
            Casilla de correo 5,
            Villa Elisa,
            Argentina. \\
       $^{5}$Radioastronomisches Institut der Universit\"at Bonn,
             Auf dem H\"ugel 71,
             53121 Bonn,
             Germany \\
}

\date{Accepted 
      Received 
      in original form }

\def\LaTeX{L\kern-.36em\raise.3ex\hbox{a}\kern-.15em
    T\kern-.1667em\lower.7ex\hbox{E}\kern-.125emX}
 
\pubyear{}

\begin{document}
\maketitle

\label{firstpage}

% abstract
%%%%%%%%%%%%%%%%
\begin{abstract}
%%%%%%%%%%%%%%%%

We compare existing high spectral resolution (R = $\lambda/\Delta\lambda \sim$ 40,000) 
Ca\,{\sc ii} K observations ($\lambda_{\rm air}$=3933.66\AA) towards 88 mainly 
B-type stars, and new observations taken using ISIS on the William Herschel Telescope 
at R $\sim$ 10,000 towards 3 stars taken from the Palomar-Green Survey, 
with 21-cm H\,{\sc i} emission-line profiles, in order to search 
for optical absorption towards known intermediate and high velocity cloud complexes. 
Given certain assumptions, limits to the gas phase abundance of Ca\,{\sc ii} 
are estimated for the cloud components. 
We use the data to derive the following distances from the Galactic plane ($z$); 
1) Tentative lower $z$-height limits of 2800 pc and 4100 pc towards 
Complex C using lack of absorption in the spectra of HD\,341617 and PG\,0855+294, respectively.  
2) A weak lower $z$-height of 1400 pc towards Complex WA-WB using lack of absorption in 
EC\,09470--1433 and a weak lower limit of 2470 pc using lack of absorption in EC\,09452--1403.
3) An upper {$z$-height} of 2470 pc towards a southern intermediate velocity cloud 
(IVC) with $v_{\rm LSR}$=--55 km\,s$^{-1}$ 
using PG\,2351+198.
4) Detection of a possible IVC in Ca\,{\sc ii} absorption at 
$v_{\rm LSR}$=+52 km\,s$^{-1}$ using EC\,20104--2944. 
No associated H\,{\sc i} in emission is detected. At this position, normal Galactic rotation 
predicts velocities of up to $\sim$ +25 km\,s$^{-1}$. The detection puts an upper $z$-height 
of 1860 pc to the cloud. 
5) Tentative H\,{\sc i} and Ca\,{\sc ii} K detections towards an IVC at $\sim$ +70 km\,s$^{-1}$ in the direction of HVC 
Complex WE, sightline EC\,06387--8045, indicating that the IVC may be at a $z$-height lower than 1770 pc. 
6) Detection of Ca\,{\sc ii} K absorption in the spectrum of PG\,0855+294 in the direction of IV20, 
indicating that this IVC has a $z$-height smaller than 4100 pc. 
7) A weak lower $z$-height of 4300 pc towards a small HVC with $v_{\rm LSR}$=+115 km\,s$^{-1}$ 
at $l,b$=200$^{\circ}$,+52$^{\circ}$, using lack of absorption in the Ca\,{\sc ii} K spectrum of PG\,0955+291.

%%%%%%%%%%%%%%
\end{abstract}
%%%%%%%%%%%%%%

\begin{keywords}
 ISM: general --
 ISM: clouds --
 ISM: structure --
 stars: early-type
\end{keywords}

%%%%%%%%%%%%%%%%%%%%%%
\section{Introduction}
%%%%%%%%%%%%%%%%%%%%%%

This paper is the second of a pair that uses a sample of mainly B-type stars to 
probe the interstellar medium of the disc and halo of the Milky Way in Ca\,{\sc ii} K 
($\lambda_{\rm air}$=3933.663 \AA). In the first (Smoker et al. 2003; hereafter Paper 1), 
we considered the abundance of Ca\,{\sc ii} K, variations in the element over degree scales, 
and the distribution of this species as a function
of distance from the Galactic plane ($z$). 
In the current work, we use the 88 sightlines in our 
sample, plus new observations towards three other stars, to search for Ca\,{\sc ii} K 
absorption in gas in intermediate and high-velocity clouds, the purpose being to try 
and improve the distance limits to these still enigmatic objects. 

Intermediate and high velocity clouds (hereafter IHVCs) are objects with absolute values of
their velocities in the local standard of rest (LSR) of between 
$\sim$40--100 and $>$ $\sim$ 100 km\,s$^{-1}$, respectively. These velocities are not explicable by simple 
rotation of gas around the Galactic centre, and hence it has been postulated that the clouds are 
either material within the Galactic halo at distances of $\sim$1--5 kpc (e.g. review by Wakker 
\& van Woerden 1997; Putman et al. 2003), or objects left over from the formation of the 
Milky Way, with distances of several hundreds of kpc (Blitz et al. 1999, Braun \& Burton 1999).  
Both types of object have been extensively studied in H\,{\sc i}, 
although until recently, the ionised component of the clouds remained 
uncertain. This situation has been rectified by results from the Wisconsin H$\alpha$ mapper (WHAM) that 
indicate that IHVCs contain ionised gas (e.g. Haffner, Reynolds \& Tufte 2001; 
Smoker et al. 2002; Tufte et al. 2002), the ionisation being caused by either collisional ionisation, 
photoionisation by the extragalactic ionising field, and/or the escape of photons from the disc of 
the Galaxy. To attempt to determine which is the most likely source of ionisation, the distance to 
IHVCs would be of great help, as to the present-day, there is still a dearth of distance measurements, 
particularly towards HVCs (Wakker 2001). The current paper once 
more attempts to address this issue, by searching for IHVC components in the Ca\,{\sc ii} 
line of a sample of mainly early-type stars, located in the Galactic disc and halo. 
The description of the reduction and analysis of the majority (88) of these stars 
was described in Paper 1, with new observations towards a further 3 objects 
being described in this paper. 

Section \ref{observations} describes new William Herschel Telescope observations towards 
three stars within IVC Complex K, Section \ref{results} presents the results of the WHT 
observations, plus a Table comparing the 91 stars in the current sample with IVC and HVC emission-line 
features found in either the Leiden-Dwingeloo Northern H\,{\sc i} survey (Hartmann \& Burton 1997), 
or the Villa-Elisa Southern H\,{\sc i} survey (Arnal et. al 2000). 

In Section \ref{disc} we discuss the Ca\,{\sc ii}\,K to H\,{\sc i} ratio, or upper limit, for 
sightlines with a IHVC H\,{\sc i} detection. In Section \ref{distance} we use these limits to 
derive new upper or lower distance estimates towards IHVCs. Finally, Section \ref{concl} contains 
the summary.

%%%%%%%%%%%%%%%%%%%%%%%%%%%%%%%%%%%%%%%%%
\section{Observations and data reduction}
\label{observations}
%%%%%%%%%%%%%%%%%%%%%%%%%%%%%%%%%%%%%%%%%

\label{newobsdist}

The new observations described in this paper were taken using the Intermediate dispersion Spectrograph 
and Imaging System (ISIS), located on the WHT, during 3--4 Aug. 2001. The
blue-arm was used, with the H2400B grating and a 1.0 arcsec slit, giving an instrumental FWHM 
resolution of $\sim$ 30 km\,s$^{-1}$ and wavelength coverage from $\sim$ 3800$-$4160 \AA. 
Three stars towards IVC Complex K were observed, PG numbers 1718+519, 1725+252 and 1738+505, which were 
reduced using standard methods to obtain the equivalent widths (EWs) and velocity centroids of the 
Ca\,{\sc ii} K components. The signal to noise ratio obtained towards the 
three objects listed above was 
$\sim$ 70, 130 and 110, respectively.

%%%%%%%%%%%%%%%%%%%%%%%%%%%%%%%%%%%%%%%%%
\section{Results}
\label{results}
%%%%%%%%%%%%%%%%%%%%%%%%%%%%%%%%%%%%%%%%%

\subsection{New WHT results}

\begin{figure}
%\epsfxsize=4.2 truecm
% %%BoundingBox: 22 360 254 636
\includegraphics{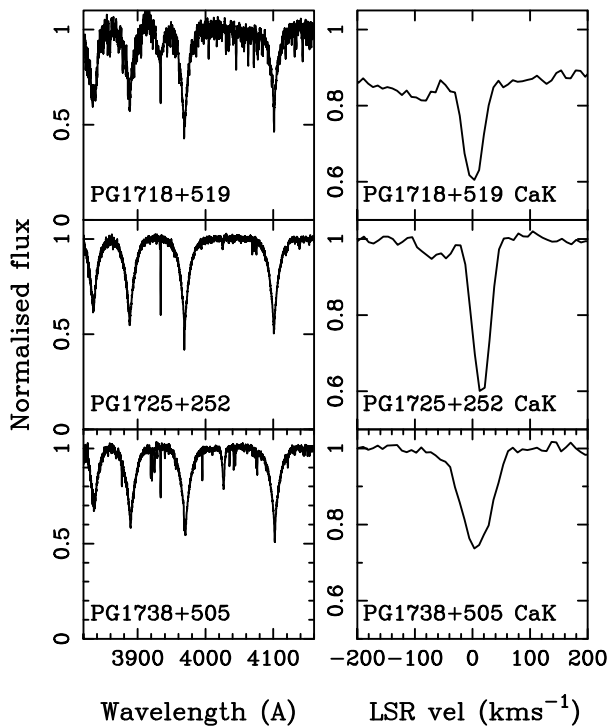}
\caption{WHT spectra of three Palomar-Green stars in the direction of Complex K. 
The left-hand panels show the entire $\lambda$-range observed, with the right-hand 
panels showing the spectra at the wavelength of Ca\,{\sc ii} K, in velocity space and 
in the Local Standard of Rest.}
\label{fig1}
\end{figure}

Fig. \ref{fig1} shows both the complete wavelength coverage, and the Ca\,{\sc ii} K line only, towards 
the three PG stars. Table \ref{tab1} shows the quantities derived from these spectra.
The Ca\,{\sc ii} K interstellar reduced equivalent width of low-velocity material towards the three sample stars 
(=$EW\times$sin($b$)) is 90, 96 and 119 m\AA. Given a Ca\,{\sc ii} K scaleheight of $\sim$ 800 pc and 
REW at infinity of $\sim$115 m\AA \, (Paper 1), this indicates that the stars are quite distant. 
Of the three objects, PG\,1718+519 turned out to be a binary, whose spectrum 
is contaminated by a late-type companion star. PG\,1725+252 is an early-type star with a 
stellar velocity of $-62\pm$6 km\,s$^{-1}$. The `intermediate velocity' line at $-63$ km\,s$^{-1}$ is hence
likely to be stellar. This was checked by running a model atmosphere code using a solar Calcium abundance with 
T$_{\rm eff}$=26,000 K, log$(g)$=5.0 (Theissen et al. 1993) and microturbulance velocity of 0 and 
5 km\,s$^{-1}$. These produced equivalent width estimates of 44 and 65 m\AA, similar to the measured value 
of 40$\pm$5 m\AA. Note that, at this high gravity a microturbulance velocity of 0 km\,s$^{-1}$ may well
be the best choice. Finally, PG\,1738+505 has a stellar velocity of 27$\pm$3 km\,s$^{-1}$ and displays 
a relatively broad interstellar profile. 

\begin{table*}
\begin{center}
\small
\caption{WHT ISIS results, with instrumental resolution $\sim$ 30 km\,s$^{-1}$, for three 
sightlines towards Complex K. PG\,1718+518 is a binary system. Limiting values for equivalent 
widths have been calculated assuming an intrinsic linewidth of 10 km\,s$^{-1}$. Values for 
the IVC H\,{\sc i} column density have been taken from the maps of Wakker (2001) and are 
between LSR velocities of --95 and --60 km\,s$^{-1}$. LV and IV 
refer to low and intermediate-velocity gas, respectively. EW refers to the equivalent width 
of Ca\,{\sc ii} K, $v$ to the velocity and FWHM to the full width half maximum of the line, 
uncorrected for instrumental effects. Where `stellar' appears in a column 
this indicates that the line is likely to be stellar in nature.
\label{tab1}
}
\begin{tabular}{lrrr}
\hline
Star                                    & PG\,1718+519          &   PG\,1725+252       & PG\,1738+505   \\
                                        &                       &                      &                \\
$l,b$ (deg)                             & 79.00, 34.94          & 48.21, 28.74         &  77.54,31.84   \\
$z$ (pc)                                & 2260                  & 320                  & 510            \\
$v_{*}^{\rm LSR}$ (km\,s$^{-1}$)        & $-$41$\pm$8           & $-62\pm6$            &    +27$\pm$3   \\
EW(Ca\,{\sc ii} K) LV (m\AA)            & 157$\pm$6             & 199$\pm$5            &    225$\pm$5   \\
EW(Ca\,{\sc ii} K)$\times$sin($b$) LV   &  90$\pm$3             &  96$\pm$3            &    119$\pm$3   \\
$v$(Ca\,{\sc ii} K) LV (km\,s$^{-1}$)   &  +1$\pm$1             & +15$\pm$1            & +6.5$\pm$1.0   \\
FWHM(Ca\,{\sc ii} K) LV (km\,s$^{-1}$)  &  37$\pm$2             & 34$\pm$1             &     61$\pm$2   \\
$v$(Ca\,{\sc ii} K) IV (km\,s$^{-1}$)   & $-90\pm 10$ (stellar) & $-63\pm3$ (stellar)  &           --   \\
FWHM(Ca\,{\sc ii} K) IV (km\,s$^{-1}$)  &  37$\pm$10            & 64$\pm$10            &           --   \\
EW(Ca\,{\sc ii} K) IV (m\AA)            &  20$\pm$7             & 40$\pm$5             &        $<$10   \\
$N$(H\,{\sc i}) IV (cm$^{-2}$)          &   10$^{19}$           & 10$^{19}$            &    10$^{19}$   \\
\hline
\end{tabular}
\normalsize
\end{center}
\end{table*}

% The upper limit to the Ca\,{\sc ii} equivalent width is $\sim$ 10 m\AA, estimated by 
% measuring the widths of weak stellar lines of known rest-wavelength. Assuming that any 
% Ca\,{\sc ii} K component is unresolved in velocity, this corresponds to a column density  
% limit of log($N$(Ca\,{\sc ii}) cm$^{-2}$)$\sim$11.0. Towards this sightline the 
% H\,{\sc i} column density from the compilation of Wakker (2001) is $\sim$ 10$^{19}$ cm$^{-2}$, integrated 
% from --60 to --95 km\,s$^{-1}$. If we assume that H\,{\sc i} gas is in a layer with exponential 
% scaleheight of 500 pc (Sembach \& Danks 1994), and given that the star is at a $z$-height of $\sim$ 510 pc 
% (Theissen et al. 1993), at this point we can predict the H\,{\sc i} column density up to the 
% star to be $\sim$ 6$\times$10$^{18}$ cm$^{-2}$, hence we estimate an upper limit to the ionic 
% gas phase abundance relative to H\,{\sc i} 
% (log$A_{\rm CaII}^{\rm IHVC}$=log($N$(Ca\,{\sc ii})--log($N$(H\,{\sc i})))=--7.7.
% This compares with the predicted value at this $N$(H\,{\sc i}) value of --7.2 (Wakker \& Mathis 2000) 
% which has an rms scatter of 0.4. Hence at the 1$\sigma$ level only, the current result implies that 
% the distance to Complex K exceeds 510 pc, although clearly this is a tentative result, given 
% the 1$\sigma$ nature of the result, uncertainties in ionisation and and the unknown 
% H\,{\sc i} column density on small scales. 
%
% for the plot_hi_Cak_together use a x-margin of 0.04 and y margin of 0.08
% with a distance between the plots of 0.02
%
\subsection{Comparison of all Ca\,{\sc ii} sample spectra with H\,{\sc i} survey data to search for IHVCs}
\label{comp1}

Table \ref{tab2} lists the sightlines from the current sample where an IHVC was detected either in 
H\,{\sc i} or Ca\,{\sc ii} K, and compares these sightlines with known 
intermediate and high-velocity cloud complexes, taken from Wakker (2001). This table 
should be used in conjunction with the notes on individual sightlines given in Sect. \ref{individual}. 
The H\,{\sc i} data are from either Leiden-Dwingeloo Northern H\,{\sc i} 
survey (Hartmann \& Burton 1997), or the Villa-Elisa Southern H\,{\sc i} survey 
(Arnal et al. 2000) for IHVCs. Columns 1 to 4 give the stellar name, 
LSR velocity in km\,s$^{-1}$, $z$-height in pc and signal-to-noise 
ratio in the stellar continuum at $\sim$ 3933 \AA,
respectively. The references for 71 of the distances are taken from Paper 1, while 
three distance estimates for the three PG stars towards the newly-observed Complex K 
are taken from Theissen et al. (1993) and de Boer et al (1997). Three other distance 
estimates are taken from Lynn et al. (2004). The remaining objects do not yet have 
distances available.

\begin{table*}
\begin{center}
\small
\caption{Comparison of H\,{\sc i} emission-line and Ca\,{\sc ii} K absorption line data 
for sightlines with a possible IHVC detection. 
The H\,{\sc i} values come either from the Leiden-Dwingeloo or Villa-Elissa surveys.
The meaning of the columns is explained in Sect. \ref{comp1}. This table should be used 
in conjunction with the notes on individual stars given in Sect. \ref{individual}. 
\label{tab2}
%
%SNR VLT stars are before smoothing
%
}
\label{obstab}
\begin{tabular}{lrrrrrrrrrrr}
\hline
    Star         & $v_{\rm LSR}^{*}$& $|z^{*}|$    
                                           & SNR   & IHVC          & $v_{\rm LSR}^{\rm IHVC}$  & FWHM  & log$N$  & $T_{B}^{\rm peak}$     &  $v_{\rm LSR}^{\rm IHVC}$  &  log$N$         & log($N_{\rm pred}$)    \\
                 &              &          & & name         & (H\,{\sc i})                       & (H\,{\sc i})   &  (H\,{\sc i})        &      (H\,{\sc i})            &  (Ca\,{\sc ii})                  &   (Ca\,{\sc ii})  &  (Ca\,{\sc ii})     \\		
                 &              &          &       &               &                &                &                &               &             &                       \\     
PG\,0009+036     &  +160        & 9090     &  35   &     PPA       & --48$\pm$5.0   &  20.0$\pm$5.0  & 19.32$\pm$0.02 & 0.59$\pm$0.02 &    --43.3   &    11.96 &     11.67  \\
EC\,00179--6503  &   +40        & 3150     &  40   &      MS       & +141.1$\pm$1.2 &  70.0$\pm$3.0  & 19.32$\pm$0.03 & 0.16$\pm$0.02 &      --     & $<$11.16 &     11.68  \\
EC\,00237--2317  &   +86        &   --     &  60   &      MS       &--121.3$\pm$1.7 &  29.4$\pm$3.5  & 19.11$\pm$0.06 & 0.22$\pm$0.02 &      --     & $<$10.98 &     11.60  \\
EC\,00321--6320  &  --30        &  664     & 120   &      MS       & +78.4$\pm$1.8  &  25.8$\pm$2.4  & 19.16$\pm$0.04 & 0.29$\pm$0.02 &      --     & $<$10.68 &     11.62  \\
        ``       &    ``        &   ``     &  ``   &      MS       &+109.8$\pm$2.4  &  27.6$\pm$4.2  & 19.10$\pm$0.04 & 0.24$\pm$0.02 &      --     & $<$10.68 &     11.60  \\
        ``       &    ``        &   ``     &  ``   &      MS       &+171.4$\pm$0.4  &   7.4$\pm$1.0  & 18.90$\pm$0.05 & 0.59$\pm$0.05 &      --     & $<$10.68 &     11.54  \\
        ``       &    ``        &   ``     &  ``   &      MS       &+180.6$\pm$0.6  &   7.5$\pm$1.1  & 18.78$\pm$0.07 & 0.42$\pm$0.05 &      --     & $<$10.68 &     11.50  \\
HD\,38666        &   +93        &  181     & 300   &   Other       &  41.0$\pm$2.0  &  11.1$\pm$2.2  & 18.71$\pm$0.02 & 0.24$\pm$0.07 &    +42.6    &    10.77 &     11.48  \\
    ``           &   +93        &   ``     &  ``   &   Other       &  52.1$\pm$0.4  &  20.7$\pm$2.8  & 19.26$\pm$0.02 & 0.44$\pm$0.03 &      --     & $<$10.28 &     11.76  \\
%    ``           &    ``        &   ``     &  ``  &   Other       & 101.5$\pm$1.2  &   4.9$\pm$2.7  & 18.11$\pm$0.16 & 0.14$\pm$0.06 &      --     & $<$10.28 &     11.30  \\
EC\,05490--4510  &   +16        &  800     &  50   &   Other       &  +50 to +80    &       --       &        --      &       --      &      --     & $<$11.06 &     --     \\
EC\,06012--7810  &   +30        & 1500     &  30   &     MB        &+207.9$\pm$1.0  &  26.9$\pm$1.7  & 18.81$\pm$0.03 & 0.13$\pm$0.02 &      --     & $<$11.28 &     --     \\
      ``         &    ``        &   ``     &       &     MB        &+288.0$\pm$0.3  &  30.8$\pm$0.6  & 19.45$\pm$0.03 & 0.43$\pm$0.07 &      --     & $<$11.28 &     11.70  \\
EC\,06387--8045  &   +49        & 1770     &  12   &     Other     &--39.4$\pm$3.5  &  56.3$\pm$5.3  & 19.43$\pm$0.04 & 0.18$\pm$0.03 &      --     & $<$11.68 &     11.70  \\
      ``         &    ``        &   ``     &       &     Other     & +70.8$\pm$1.0  &  19.6$\pm$2.0  & 18.82$\pm$0.10 & 0.17$\pm$0.02 &   +75.0     &    11.50 &     11.52  \\
PG\,0833+699     &   +23        & 1980     &  30   &     LLIV Ar   &--69.4$\pm$0.6  &  15.7$\pm$1.3  & 19.34$\pm$0.04 & 0.71$\pm$0.04 &  --67.0     &    11.46 &     11.67  \\
      ``         &              &   ``     &       &     LLIV Ar   &--40.1$\pm$0.5  &  26.9$\pm$1.5  & 19.78$\pm$0.03 & 1.17$\pm$0.04 &  --45.0     &    11.28 &     11.80  \\
PG\,0855+294     &   +58        & 4100     &  40   &       C       &--168.7$\pm$1.3 &  22.8$\pm$3.5  & 19.00$\pm$0.04 & 0.23$\pm$0.03 &      --     & $<$11.16 &     11.57  \\
      ``         &    ``        &   ``     &  ``   & IV Arch       & --23.1$\pm$1.0 &   4.8$\pm$0.5  & 19.44$\pm$0.10 & 3.20$\pm$0.05 &      --     &       -- &        --  \\
      ``         &    ``        &   ``     &  ``   & IV Arch       & --27.0$\pm$3.0 &  22.0$\pm$2.0  & 20.07$\pm$0.10 & 2.47$\pm$0.10 &  --29.0     &    11.90 &     11.89  \\
EC\,09452--1403  &  +226        & 2470     &  25   &      WB       & +120.0$\pm$10  &  24.6$\pm$10.2 & 18.65$\pm$0.16 & 0.10$\pm$0.04 &      --     & $<$11.36 &     11.46  \\
EC\,09470--1433  &    --        & 1400     &  60   &     Other     &  +55.0$\pm$5.0 &  20.0$\pm$5.0  & 18.95$\pm$0.09 & 0.25$\pm$0.03 &      --     & $<$10.98 &     11.35  \\
      ``         &    ``        &   ``     &  ``   &      WB       & +114.0$\pm$8.0 &  25.0$\pm$5.0  & 18.97$\pm$0.05 & 0.25$\pm$0.03 &      --     & $<$10.98 &     11.35  \\
PG\,0955+291     &   +72        & 4300     &  35   &     IV Spur   & --29.3$\pm$1.2 &  59.2$\pm$2.0  & 19.81$\pm$0.02 & 0.58$\pm$0.02 &      --     & $<$11.22 &     11.81  \\
      ``         &    ``        &   ``     &  ``   &      ``       &      --        &      --        &      --        &      --       &  --45.0     &    11.74 &    --      \\
      ``         &    ``        &   ``     &  ``   &      ``       &      --        &      --        &      --        &      --       &  --64.0     &    11.26 &    --      \\
      ``         &    ``        &   ``     &  ``   &     Other     & +115.5$\pm$2.3 &  20.9$\pm$4.0  & 18.60$\pm$0.10 & 0.09$\pm$0.03 &      --     & $<$11.22 &     11.45  \\
PG\,1008+689     &  --11        &  950     &  45   & IV Arch       & --44.3$\pm$0.3 &  21.9$\pm$0.8  & 19.62$\pm$0.03 & 1.30$\pm$0.06 &      --     & $<$11.11 &    --      \\
      ``         &    ``        &   ``     &  ``   & IV Arch       &      --        &       --       &        --      &       --      &   --39.0    &    11.48 &    --      \\
      ``         &    ``        &   ``     &  ``   & IV Arch       &      --        &       --       &        --      &       --      &   --49.0    &    11.62 &    --      \\
EC\,10087--1411  &   +96        &  620     &  40   &     Other     &  +63.0$\pm$5   &  32.2$\pm$4.6  & 19.30$\pm$0.04 & 0.32$\pm$0.04 &      --     & $<$11.16 &     11.66  \\
EC\,11074--2912  &    --        &  950     &  25   &     Other     &--48.6$\pm$3.3  &  36.5$\pm$10.5 & 19.36$\pm$0.05 & 0.32$\pm$0.04 &      --     & $<$11.36 &    --      \\
EC\,11507--2253  &  +210        &   --     &  45   &     Other     & +45.0$\pm$10   &  30.0$\pm$10   & 19.11$\pm$0.06 & 0.22$\pm$0.03 &      --     & $<$11.11 &    --      \\
PG\,1213+456     &  --15        & 2700     &  20   & IV Arch       &--57.8$\pm$0.3  &  22.8$\pm$0.9  & 19.55$\pm$0.02 & 0.82$\pm$0.02 &  --56.9     &    11.83 &   11.73    \\
PG\,1243+275     &  +107        & 6200     &  25   &IV Ar/Sp       &--23.7$\pm$0.4  &  39.4$\pm$0.8  & 19.91$\pm$0.01 & 1.08$\pm$0.02 &      --     & $<$11.36 &    --      \\
      ``         &    ``        &   ``     &  ``   &IV Ar/Sp       &      --        &       --       &        --      &       --      &  --42.8     &    11.28 &    --      \\
LS\,3510         &     0        &   --     & 200   &   Other       & +38.1$\pm$5.0  &  17.5$\pm$4.0  & 20.52$\pm$0.10 &       --      &   --       & $<$10.46 &    12.03     \\
    ``           &    ``        &   --     &  ``   &   Other       & +53.4$\pm$5.0  &  11.7$\pm$4.0  & 20.24$\pm$0.10 &       --      &   --       & $<$10.46 &    11.94     \\
LS\,3604         &  --30        &   --     & 160   &   Other       & +38.8$\pm$1.0  &   37.8$\pm$2.5 & 21.13$\pm$0.03 &       --      &   --       & $<$10.56 &    12.21     \\
    ``           &    ``        &   ``     & ``    &   Other       & +41.8$\pm$0.6  &    7.6$\pm$0.5 & 20.31$\pm$0.03 &       --      &   --       & $<$10.56 &    11.96     \\
    ``           &    ``        &   ``     & ``    &   Other       & +51.0$\pm$0.5  &    4.0$\pm$0.5 & 19.63$\pm$0.03 &       --      &   --       & $<$10.56 &    11.76     \\
LS\,3694         &  --29        &   --     & 150   &   Other       & +40.9$\pm$5.0  &  44.1$\pm$5.2  & 20.86$\pm$0.04 &       --      &   --       & $<$10.58 &    12.13   \\
    ``           &  ``          &   --     & ``    &   Other       & +55.6$\pm$1.0  &   7.8$\pm$1.0  & 20.06$\pm$0.04 &       --      &   --       & $<$10.58 &    11.89   \\
LS\,3751         &  --24        &   --     & 170   &   Other       &     --         &      --        &      --        &       --      &  +38.6     &    11.33 &     --     \\
    ``           &   ``         &   --     & ``    &   Other       & +56.9$\pm$0.5  &   5.8$\pm$0.8  & 19.86$\pm$0.03 &       --      &   --       & $<$10.52 &    11.83   \\
    ``           &   ``         &   --     &  ``   &   Other       & +59.7$\pm$0.5  &  18.5$\pm$2.0  & 20.16$\pm$0.03 &       --      &   --       & $<$10.52 &    11.92   \\
PG\,1725+252     &  --64        &  320     & 120   &       C       &--154.0$\pm$1.7 &  29.8$\pm$3.6  & 19.47$\pm$0.03 & 0.36$\pm$0.04 &   --       & $<$11.28 &     11.71  \\
PG\,1738+505     &   +24        &  510     & 100   &       K       & --98.2$\pm$0.7 &  26.4$\pm$2.1  & 19.47$\pm$0.03 & 0.59$\pm$0.04 &   --       & $<$11.36 &     11.71  \\
HD\,341617       &   +63        & 2800     & 110   &       C       &--118.0$\pm$10  &  35.0$\pm$10.0 & 19.18$\pm$0.04 & 0.31$\pm$0.02 &   --       & $<$10.62 &     11.62  \\
     ``          &    ``        &   ``     &  ``   &  Other        &      --        &       --       &                &       --      &  --45.0    &    10.80 &     --     \\
NGC6712 ZNG-1    & --102        &  500     &  50   &    MS         &      --        &       --       &        --      &       --      &   --       & $<$11.06 &     --     \\
LS\,5112         & --120        &  --      & 150   &  GCN?, MS     &      --        &       --       &        --      &       --      &   --137.3  &    11.33 &     --     \\
% M22 ZNG-15     & --           &  330     & 250   &    --         &      --        &       --       &        --      &       --      &   --       & $<$10.36 &     --     \\
EC\,19071--7643  &  --17        &  970     & 110   &     Other?    &--40 to --60    &       --       &        --      &       --      &   --       & $<$10.72 &     --     \\
EC\,19489--5641  &    --        &   --     &  60   &     Other     &+36.0$\pm$3.0   &  15.0$\pm$5.0  & 19.35$\pm$0.05 & 0.68$\pm$0.03 &   --       & $<$10.98 &     11.68  \\
EC\,19490--7708  &    --        &  800     &  70   &     Other     &+43.0$\pm$1.4   &  23.8$\pm$2.8  & 19.08$\pm$0.06 & 0.30$\pm$0.04 &   --       & $<$10.91 &     11.59  \\
EC\,19596--5356  &  +200        &20000     &  55   &     Other     &+38.7$\pm$0.4   &  12.1$\pm$0.9  & 19.15$\pm$0.02 & 0.58$\pm$0.04 &  +43.4     &  11.45   &    11.61   \\
EC\,20089--5659  &  --17        &  409     & 140   &     Other     & +50.0$\pm$1.0  &  28.0$\pm$1.7  & 19.50$\pm$0.03 & 0.57$\pm$0.03 &   --       & $<$10.61 &     11.72  \\
EC\,20104--2944  &  +145        & 1860     &  70   &     Other     & +44$\pm$5      &  34.4$\pm$2.6  & 19.30$\pm$0.04 & 0.30$\pm$0.02 &  +52.7     &  11.72   &    11.66   \\
M\,15\,ZNG--1    & --100        & 4600     &  80   &      gp       & +69.6$\pm$0.6  &  18.2$\pm$1.4  & 19.34$\pm$0.04 & 0.61$\pm$0.04 &  +65.0     &  12.38   &   11.67    \\
EC\,23169--2235  &   +79        & 2220     &  30   &     Other     & --30 to --50   &       --       &        --      &       --      &   --       & $<$11.28 &     --     \\
PG\,2351+198     & --275        & 2470     &  50   &    IVS        &--55.0$\pm$5.0  &  20.0$\pm$5    & 18.70$\pm$0.14 & 0.14$\pm$0.05 & --55.80    &  11.12   &   11.48    \\
\hline						       			                            
\end{tabular}     			                            
\end{center} 			                            
\normalsize
\end{table*}

\begin{table}
\begin{center}
\small
\caption{Sightlines with neither a H\,{\sc i} nor Ca\,{\sc ii} IHVC detection. 
The meaning of the columns is explained in Sect. \ref{comp1}. 
}
\label{tab3}
\begin{tabular}{lrrrr}
\hline
    Star         & $v_{\rm LSR}^{*}$& $|z^{*}|$    
                                           & SNR    &  log$N$         \\
                 &              &          &        &  (Ca\,{\sc ii}) \\		
                 &              &          &        &                 \\     
EC\,00358--1516  &   +84        & 4294     & 120    & $<$10.68   \\
EC\,00468--5622  &    +4        & 1971     & 140    & $<$10.61   \\
EC\,01483--6806  &   +61        & 2100     & 100    & $<$10.76   \\
EC\,03240--6229  &  --17        & 1600     &  30    & $<$11.28   \\
EC\,03342--5243  &   +84        & 1210     & 200    & $<$10.46   \\
EC\,03462--5813  &   +24        &  650     &  50    & $<$11.06   \\
EC\,04420--1908  &  +192        & 1124     &  35    & $<$11.22   \\
EC\,04460--3215  &  --17        &  880     &  20    & $<$11.46   \\
EC\,05229--6058  &   +17        & 1060     &  30    & $<$11.28   \\
EC\,05438--4741  &   +37        & 1800     &  35    & $<$11.22   \\
EC\,05515--6107  &   +75        & 2280     &  35    & $<$11.22   \\
EC\,05515--6231  &  --21        &  335     & 100    & $<$10.76   \\
EC\,05582--5816  &   +66        &  670     &  45    & $<$11.11   \\
PG\,0823+499     &   +12        & 1000     &  45    & $<$11.11   \\
PG\,0914+001     &   +80        & 8440     &  35    & $<$11.22   \\
PG\,0934+145     &  +105        & 5820     &  30    & $<$11.28   \\
EC\,09414--1325  &   +60        & 1540     &  40    & $<$11.16   \\
PG\,0954+049     &   +90        & 2400     &  30    & $<$11.28   \\
EC\,10500--1358  &   +92        & 3330     &  30    & $<$11.28   \\
EC\,10549--2953  &  --15        &  800     &  40    & $<$11.16   \\
HD\,97917        &   +11        &   --     & 120    & $<$10.68   \\ 
PG\,1205+228     &  +156        & 2340     &  60    & $<$10.98   \\
PG\,1212+369     &  --32        & 2600     &  30    & $<$11.28   \\
PG\,1310+316     &  --55        & 8100     &  20    & $<$11.45   \\
EC\,13139--1851  &   +18        & 1060     &  30    & $<$11.28   \\
PG\,1323--086    &  --41        &12600     & 140    & $<$10.61   \\
PG\,1351+393     &  --24        & 6400     &  25    & $<$11.36   \\
EC\,14102--1337  &  --20        &   --     &  40    & $<$11.16   \\
HD\,137569       &  --24        &  500     & 400    & $<$10.16   \\
EC\,15374--1552  &  --56        &   --     &  50    & $<$11.06   \\
LS\,IV--0401     &  +105        & 4700     &  80    & $<$10.86   \\
M\,10\,ZNG--1    &   +90        & 2970     & 140    & $<$10.61   \\
PG\,1704+222     &  --22        & 3700     &  80    & $<$10.85   \\
PG\,1708+142     &  +180        &10000     &  30    & $<$11.28   \\
PG\,1718+519     &  --41        & 2260     & 100    & $<$10.76   \\
M\,22\,ZNG-5     & --130        &  330     &  70    & $<$10.92   \\
EC\,19304--5337  &  +165        &   --     &  60    & $<$10.98   \\
EC\,19337--6743  &   --8        &  390     & 150    & $<$10.58   \\
EC\,19476--4109  &   --5        &  842     & 110    & $<$10.72   \\
EC\,19563--7205  &  --10        &   --     &  15    & $<$11.58   \\
EC\,19579--4259  &   +16        &  180     & 120    & $<$10.68   \\
EC\,19586--3823  &  --96        & 1513     &  70    & $<$10.92   \\
EC\,20011--5005  & --168        & 3927     &  40    & $<$11.16   \\
EC\,20068--7324  &   +71        &   --     &  30    & $<$11.28   \\
%EC\,20140--6935 & +0 del       & 1190     & nodata &    --      \\
%EC\,20153--6731 & --39  del    & 1450     & nodata &    --      \\
EC\,20252--3137  &   +23        & 1642     & 100    & $<$10.76   \\
EC\,20292--2414  &    +8        & 1000     &  35    & $<$11.22   \\
EC\,20411--2704  &   +18        &  200     &  30    & $<$11.28   \\
EC\,20485--2420  &  --40        & 2100     & 110    & $<$10.72   \\
PG\,2120+062     &  --56        & 2500     &  50    & $<$11.06   \\
PG\,2146+087     &   +19        & 1250     &  45    & $<$11.11   \\
PG\,2219+094     &  --17        & 4190     &  60    & $<$10.98   \\
PG\,2229+099     &  --10        & 5220     &  12    & $<$11.68   \\
PG\,2345+241     &   +82        & 2920     &  60    & $<$10.98   \\
PG\,2356+167     &    +3        & 2030     &  30    & $<$11.28   \\
\hline						       			                            
\end{tabular}     			                            
\end{center} 			                            
\normalsize
\end{table}

If the sightline is towards a known IVC or HVC, Columns 5--9 give the IHVC name (mostly 
taken following Wakker 2001) LSR velocity in km\,s$^{-1}$ of H\,{\sc i} gas at this position on the sky, full 
width half maximum value (FWHM) of the H\,{\sc i} profile in km\,s$^{-1}$, 
log(H\,{\sc i} column density in cm$^{-2}$) and peak brightness temperature in K, 
respectively. These values were obtained via Gaussian profile fitting using {\sc elf} 
within {\sc dipso} (Howarth et al. 1996). In the Northern hemisphere, the H\,{\sc i} values are taken 
from the nearest stray-radiation corrected Leiden-Dwingeloo H\,{\sc i} survey pointing (Hartmann \& Burton 1997), 
which sampled the sky north of declination=$-$30 degrees at a resolution of 0.5 degrees. In the 
Southern hemisphere, the corresponding values are from the Instituto Argentino de Radioastronom\'ia Villa-Elisa 
Southern Sky survey (Arnal et al. 2000), which surveyed the sky south of declination=$-$25 deg. The version 
of the Villa-Elisa survey that we have used has been corrected for the effects of stray radiation. 
In this case, linear interpolation of the four nearest profiles was performed. 
In the region of overlap between the two surveys we used the Leiden-Dwingeloo data. 

In the case where optical absorption is detected, Column 10 gives the LSR velocity assuming that 
the feature is due to Ca\,{\sc ii} K, with Column 11 giving either the 
log(Ca\,{\sc ii} column density in cm$^{-2}$) of the IHVC, or (more frequently) its 
upper limit from the current dataset. Given the instrumental full width half maximum 
resolution, $\Delta\lambda_{\rm instr}$, the observed SNR in the continuum, $\sigma_{\rm cont}$, 
was used to calculate a limiting equivalent width in \AA, $EW_{\rm lim}$(Ca\,{\sc ii}), thus;

\begin{equation}
EW_{\rm lim}({\rm Ca \,II})=5\sigma_{\rm cont}^{-1} \Delta\lambda_{\rm instr}.
\label{eqlim}
\end{equation}

Typically, the spectra have a instrumental resolution of 0.1\AA \, and SNR=30, which means that the 
calculated limiting equivalent width is on the order of 20 m\AA, which is compatible with visual 
inspection of the spectra. Once this limiting equivalent width has been estimated, it can be used 
to determine the limiting column density $N_{\rm lim}$, assuming that we are on the linear part of 
the curve of growth, viz;

\begin{equation}
N_{\rm lim}=1.13 \times 10^{20} \frac{EW_{\rm lim}}{\lambda^{2}f},
\label{nlim}
\end{equation}

\noindent
where $\lambda$ is the wavelength in \AA \, and $f$ is the oscillator strength of 
Ca\,{\sc ii} K, taken 
to be 0.634 (Morton 1991). Finally, Column 12 gives the predicted 
value of the Ca\,{\sc ii} column density that would be expected from the observed 
H\,{\sc i} column, viz;

\begin{equation}
{\rm log}(N_{\rm p}^{\rm WM00})({\rm CaII}) = (0.30 \times ({\rm log}(N_{H})) + 5.87,
\label{wm00CaII}
\end{equation}

\noindent
which is taken from Wakker \& Mathis (2000, henceforth WM00). 
Note that using this relationship only gives an approximate estimation of the real Ca\,{\sc ii} 
column density because; 1) The H\,{\sc i} column density is derived from a large beam and does 
not take fine-scale structure into account, 2) The H-to-Ca ratio varies somewhat from cloud-to-cloud, 
3) the H-to-Ca ratio may vary within clouds and 4) We are using $N_{\rm HI}$ and not $N_{\rm Htot}$. 
Savage et al. (2000) find that $N$(H\,{\sc i})(Ly-alpha)=0.6--1.0$\times N$(H\,{\sc i})(21-cm). 

% Fig. \ref{fig2} shows the $A_{\rm CaII}^{\rm IHVC}$ vs. $N$(H\,{\sc i}) correlation plot for the current sample, 
% where the numbers refer to the individual sightlines. Also plotted on the figure is the 
% best-fit value and deviation for known IHVC complexes, taken from Wakker \& Mathis (2000). 

Finally, Table \ref{tab3} lists the sightlines where no IHVC was detected in either 
the H\,{\sc i} or optical Ca\,{\sc ii} K spectra.

%%%%%%%%%%%%%%%%%%%%%%%%%%%%%%%%%
\section{Discussion}
\label{disc}
%%%%%%%%%%%%%%%%%%%%%%%%%%%%%%%%%

In this section, we give notes on individual sightlines, then collate the information 
in Sect. \ref{distance} to try to improve the distance estimates to IHVC 
complexes.

\subsection{Notes on individual sightlines}
\label{individual}

We first provide notes on individual sightlines, comparing H\,{\sc i} data from 
either the Leiden-Dwingeloo or Villa-Elisa 21-cm H\,{\sc i} surveys with our optical Ca\,{\sc ii} 
K absorption line profiles. Where there is either H\,{\sc i} emission or Ca\,{\sc ii} K absorption 
seen with $|v_{\rm LSR}| >\sim$ 40 km\,s$^{-1}$, the sightline is discussed below. The remaining 
sightlines are not considered in the current paper. Where previous authors have determined distances 
to IVCs or HVCs using the same stars as the current dataset, the references are given. Lack of references 
hence implies that these sightlines have not previously been searched for IVCs/HVCs. For each sightline, 
we also give an estimate of the Ca\,{\sc ii} column density predicted from the H\,{\sc i} column density 
and estimated using equation \ref{wm00CaII}. The 1$\sigma$ scatter on this relation is 0.42 dex (WM00). 
The number of standard deviations that the upper limit is away from the predicted value is 
also given. If the observed upper limit to Ca\,{\sc ii} column density is significantly lower than that 
predicted by equation \ref{wm00CaII}, this implies that, {\it under the assumptions given in Sect. 
\ref{comp1}}, then a non-detection is likely to be caused by the star being closer than 
the IHVC. Note that for the current sample, we consider only the H\,{\sc i} component, and neglect any 
ionised hydrogen present. Presence of H\,{\sc ii} would increase the estimated value of 
log($N_{\rm CaII}^{\rm WM00}$) and make it more likely that a non-detection could be used to 
derive a lower distance limit. In some IVCs at least, depending on the filling factor, the ionised 
component of H could be as much as that contained in neutral material (Smoker et al. 2002). 

\begin{figure*}
\includegraphics{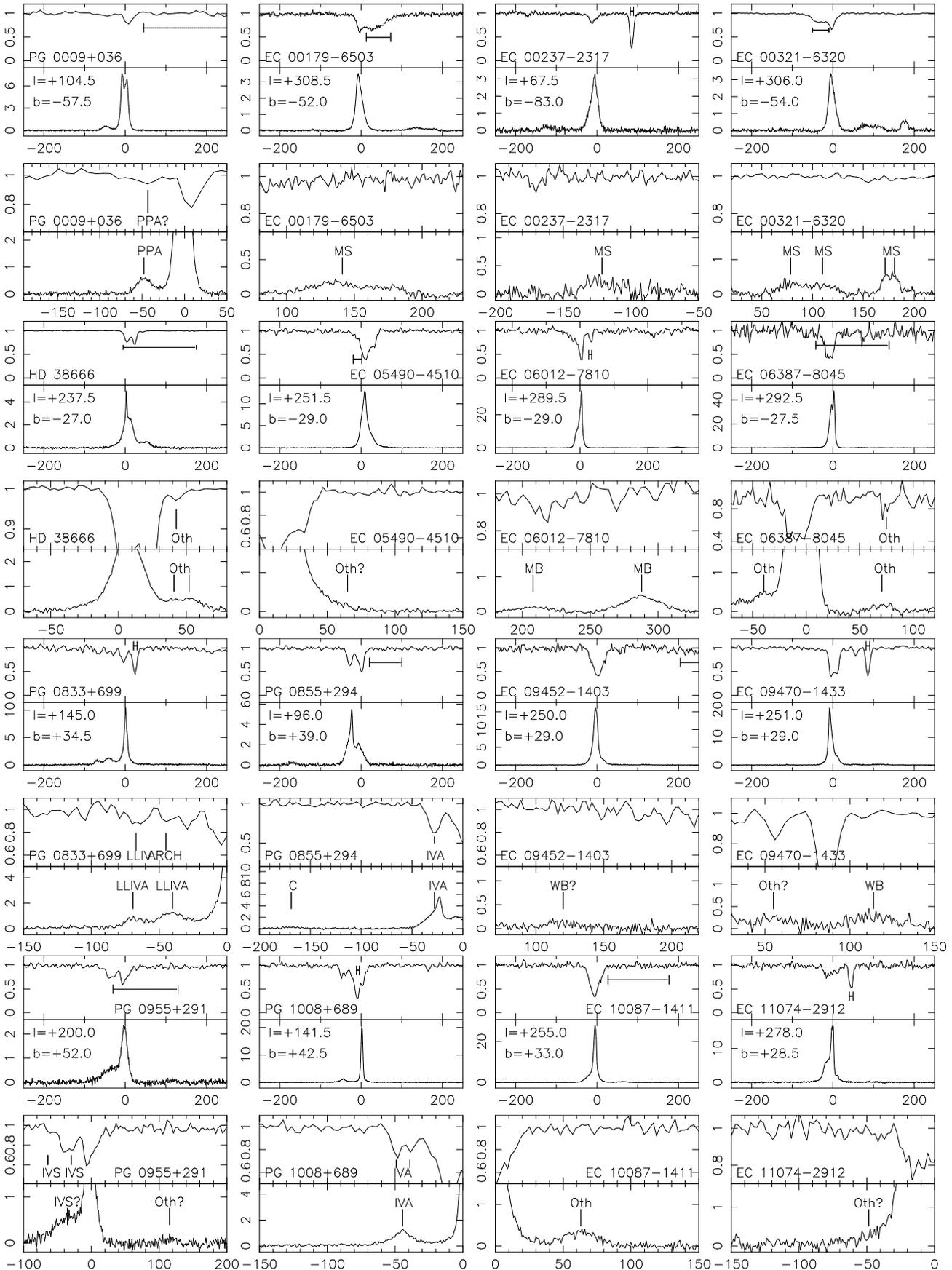}
\caption{H\,{\sc i} spectra taken from either the Leiden-Dwingeloo or Villa-Elisa 
survey with possible IVC or HVC detections, plotted beneath Ca\,{\sc ii} K spectra 
from the current sample. Two sets of H\,{\sc i}/Ca\,{\sc ii} K plots are shown per 
sightline, with different scales. On the bottom plots are marked possible IHVC detections. 
Horizontal lines indicate the extent of the stellar FWHM 
profile.}
\label{fig2}
\end{figure*}

Individual sightlines, towards which there are possible or likely IHVC detections, are now discussed.  
The H\,{\sc i} and Ca\,{\sc ii} K spectra towards 
these sightlines being displayed in Fig. \ref{fig2}. Where there exist higher-resolution 21-cm H\,{\sc i} 
spectra towards the sample stars, this is noted in the comments for the individual sightlines. 

\setcounter{figure}{1}
\begin{figure*}
\includegraphics{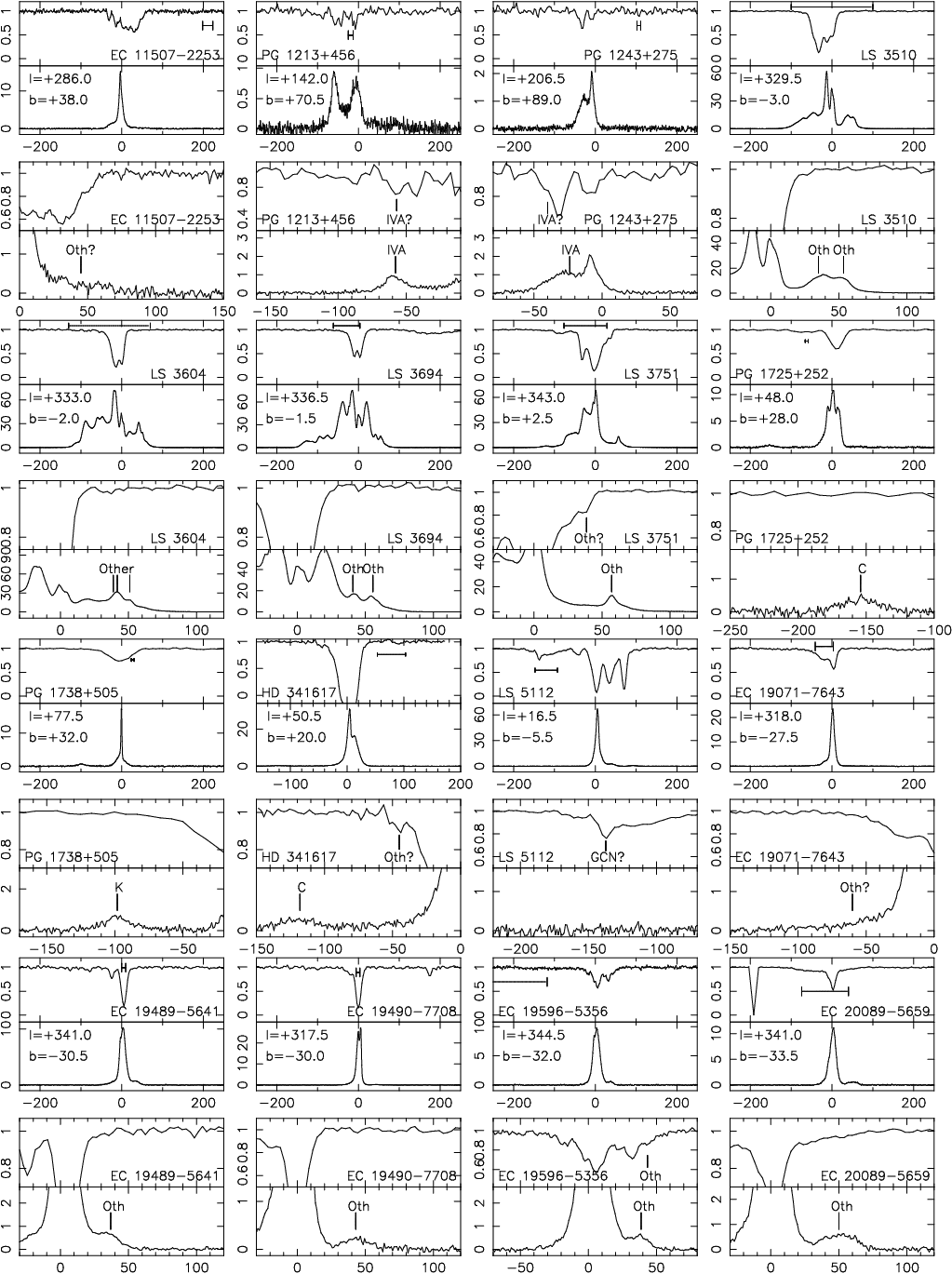}
\caption{{\it $-$ continued.}}
\end{figure*}
\setcounter{figure}{1}
\begin{figure*}
\includegraphics{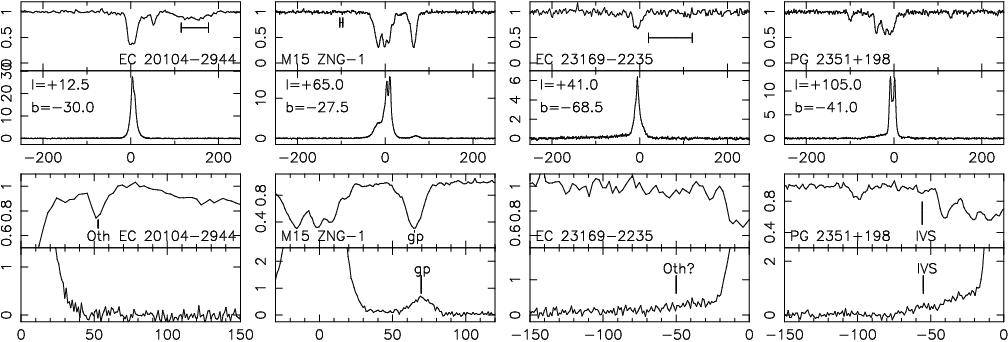}
\caption{{\it $-$ continued.}}
\end{figure*}

{\it PG\,0009+036}: The Dwingeloo H\,{\sc i} spectrum shows an IVC at --48 km\,s$^{-1}$. 
This material lies in the direction of the Pegasus-Pisces Arch (PPA; Wakker 2001).
This star was previously observed by de Boer et al. (1994) with IUE, although
the spectrum was of low quality and no limit to the IVC distance could be set. De Boer et al. 
(1994) also presented a Effelsberg H\,{\sc i} profile in this direction, which is towards 
Magellanic Stream Cloud V. They find a log($N_{\rm HI}$) value of 19.26 for the IVC,  
close to the LDS value of 19.32. Our Ca\,{\sc ii} spectrum shows {\em tentative} evidence for 
absorption at --43.4 km\,s$^{-1}$, although the signal to noise is low. If the feature
is real, then it has a log($N_{\rm CaII}$) value of 11.96, compared to the WM00 estimated 
value of 11.67. In this case, the $z$-height of the cloud would be less than 
$\sim$ 9000 pc.

{\it EC\,00179--6503}: This sightline lies in the direction of the Magellanic Stream.
The Villa-Elisa H\,{\sc i} spectrum shows HV material at $\sim$ +145 km\,s$^{-1}$, possibly 
made up of two separate components with central velocities of +135 and +177 km\,s$^{-1}$ and FWHM velocity 
widths of 48 and 21 km\,s$^{-1}$. A single-component fit has a  FWHM of $\sim$ 70 km\,s$^{-1}$. 
There is no corresponding Ca\,{\sc ii} detection to a limit of 
log($N_{\rm CaII}$)=11.16. At this H\,{\sc i} column density WM00 predict 
log($N_{\rm CaII}^{\rm WM00}$)=11.66, hence at the 1.5$\sigma$ level this HVC is likely to
be at a $z$-height exceeding $\sim$ 3150 pc. 

{\it EC\,00237--2317}: This sightline lies in the direction of the Magellanic Stream.
The  Dwingeloo H\,{\sc i} spectrum shows a HVC at --120 km\,s$^{-1}$, although the 
column density is low ($N_{\rm HI}$=1.3$\times$10$^{19}$ cm$^{-2}$). 
There is no corresponding detection in Ca\,{\sc ii} K with log($N_{\rm CaII}$)$<$ 10.98. 
For this $N_{\rm H}$, WM00 predict log($N_{\rm CaII}^{\rm WM00}$)=11.60, hence at a 1.5$\sigma$ level, 
the lack of Ca\,{\sc ii} absorption implies that the HVC is further away than the star. The distance 
towards EC\,00237--2317 is currently unknown, no Stromgren photometry exists and only the 
Si\,{\sc ii} line is observed, so no temperature has been derived for the object (Lynn, unpublished 
result). Note that the optical absorption feature at --190 km\,s$^{-1}$ is likely to be S\,{\sc ii} 
at $\lambda_{\rm air}$=3931.91\AA. 

{\it EC\,00321--6320}: The Villa-Elisa H\,{\sc i} spectrum shows an IVC and two HVCs at +78.4, +109.8 and 
+175 km\,s$^{-1}$ with log($N_{\rm HI}$) values of 19.16, 19.10 and 19.14 respectively. 
The Parkes spectrum shown in Wakker et al. (2001) (sightline HD\,003175)  has components at +73, 
+104 and +168 km\,s$^{-1}$, with log($N_{\rm HI}$) values of 19.00, 18.91 and 19.41 respectively. 
Optical absorption at +155 km\,s$^{-1}$ is present that merges in with the 175 km\,s$^{-1}$ H\,{\sc i} 
feature, but is likely to be He\,{\sc i} at $\lambda_{\rm air}$=3935.95\AA. The IVC/HVC at +78.4 and 
+109.8 km\,s$^{-1}$ are not detected in Ca\,{\sc ii} to a limit of log($N_{\rm CaII}$)=10.68, compared with the WH00 
prediction of 11.62 and 11.60, respectively. Thus at the 2.5$\sigma$ level, these HI features are likely to be further 
away than the stellar $z$-height of 664 pc. Using the H\,{\sc i} results compiled in Wakker et al. (2001) 
results in the same conclusion. 

%{\it EC\,05438--4741}: This sightline passes near to clouds of HVC complex WC. 
%The Villa-Elisa H\,{\sc i} spectrum shows a possible IVC feature at +86 km\,s$^{-1}$ 
%with a peak of only 0.24 K. Although there is also weak Ca\,{\sc ii} absorption seen at this 
%velocity, it could very well be the edge of the stellar line which lies at +37 km\,s$^{-1}$. Hence the 
%current data does not constrain the distance. 

{\it HD\,38666 ($\mu$ Col.)}: The Villa-Elisa H\,{\sc i} spectrum shows blended IVC components at $\sim$ +41
and $\sim$ +53 km\,s$^{-1}$.
%, with a marginal HVC component detected at +101.5 km\,s$^{-1}$. 
This star has been observed in the UV with the Goddard echelle spectrograph, and the results discussed 
by both Howk, Savage \& Fabian (1999) and Brandt et al. (1999). The latter found absorption features 
at +3, +21, +33 and +42 km\,s$^{-1}$. The +42 km\,s$^{-1}$ component
was also detected in Ca\,{\sc ii} K in the current spectrum. It thus seems likely that there are two IVC 
components, one at +42 km\,s$^{-1}$ and one at +52 km\,s$^{-1}$. The latter is not detected in 
Ca\,{\sc ii} K to a limit of log($N_{\rm CaII}$)=10.28, 
compared to the WM00 prediction of 11.76. This indicates that this component is likely to be further away 
than HD\,38666, which has a $z$-height of $\sim$ 181 pc.

{\it EC\,05490--4510}: The Villa-Elisa H\,{\sc i} spectrum shows a weak extended wing in H\,{\sc i} from +50 
to +80 km\,s$^{-1}$ that merges with LV gas. No such wing is present in the Ca\,{\sc ii} data although this 
is too faint to be detected in any case with the current data. 

{\it EC\,06012--7810}: This sightline lies in the direction of the Magellanic Bridge.
The Villa-Elisa H\,{\sc i} spectrum shows possible emission at +207 km\,s$^{-1}$, 
plus a separate peak at +288 km\,s$^{-1}$. WM00 predict log($N_{\rm CaII}^{\rm WM00}$)=11.70 
for the latter feature, compared to the observational limit of log($N_{\rm CaII}$)$<$11.28. Hence at the 
$\sim$ 3$\sigma$ level, the +294 km\,s$^{-1}$ feature is at a $z$-height exceeding $\sim$ 1500 pc. 
Finally, note that the optical absorption feature at +216 km\,s$^{-1}$ is likely to be He\,{\sc i} at 
$\lambda_{\rm air}$=3935.95\AA. 
 
{\it EC\,06387--8045}: The Villa-Elisa H\,{\sc i} spectrum shows a marginal H\,{\sc i} detection at 
--39 km\,s$^{-1}$ that merges in with LV gas and also at +71 km\,s$^{-1}$. The latter feature shows a 
tentative detection in Ca\,{\sc ii} K at +75 km\,s$^{-1}$, although the SNR is low. {\it If} both features 
are real and are the same parcel of gas, then this IVC has an upper $z$-height limit of 1770 pc and 
log($N_{\rm CaII}$) value of 11.50, compared with the predicted value from WM00 of 11.52.

{\it PG\,0833+699}: This sightline has been discussed previously by Ryans et al. (1997b) who used 
the current data to determine an upper limit to the distance of LLIV1. The Dwingeloo H\,{\sc i} 
spectrum shows two strong components at --69.3 and --40.1 km\,s$^{-1}$ with log($N_{\rm HI}$) values 
of 19.34 and 19.78, merged with LV gas. The Lovell-telescope H\,{\sc i} profile 
from Ryans et a. (1997b) shows the same components at --70.5 and --43.0 km\,s$^{-1}$ with log($N_{\rm HI}$) 
values of 19.36 and 20.01. As noted by Ryans (1997b), both of these components are detected 
in Ca\,{\sc ii} in absorption, giving an upper limit of z=1980 pc towards these IVCs.

\begin{figure}
\includegraphics{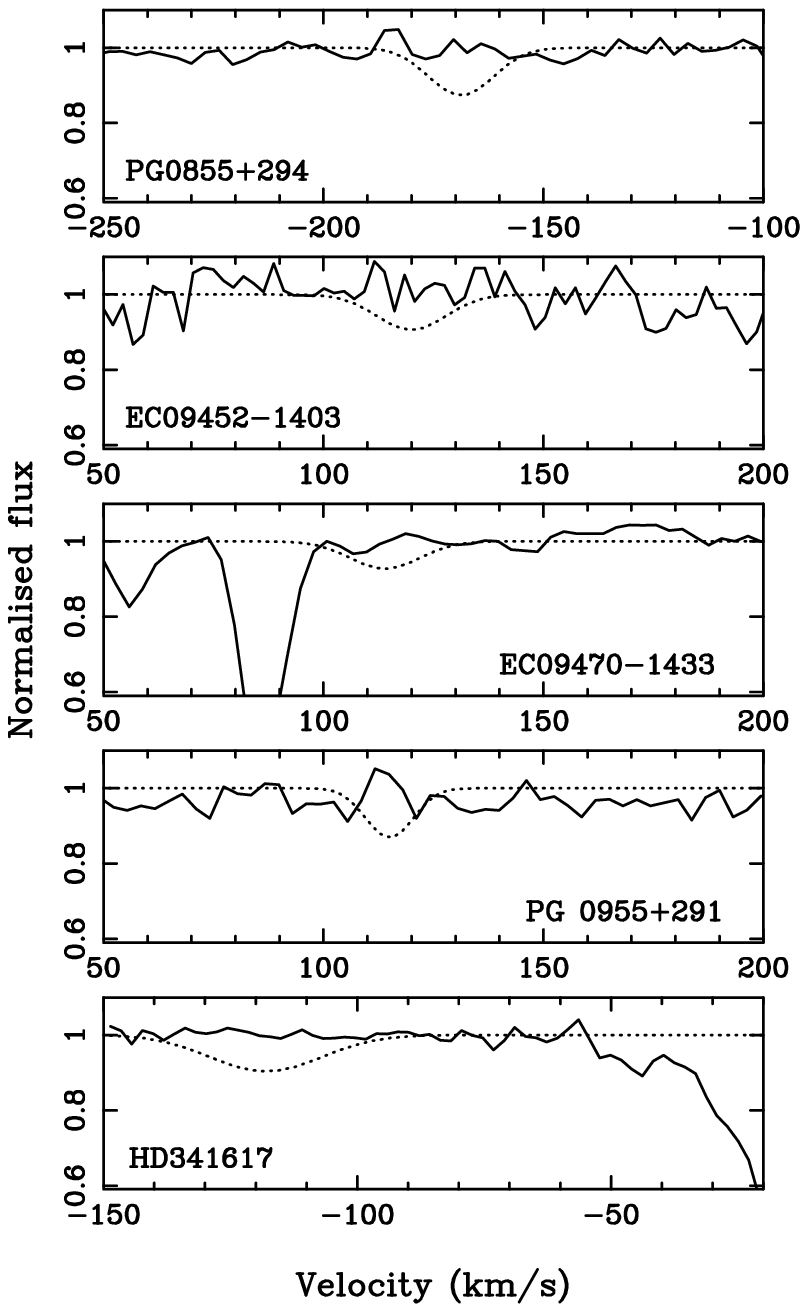}
\caption{Solid lines: observed Ca\,{\sc ii} K spectra towards four IHVCs towards which H\,{\sc i} is detected. 
Dashed lines: Ca\,{\sc ii} K model-fit spectra calculated using the $v_{\rm LSR}^{\rm IHVC}$, FWHM(H\,{\sc i}) and log($N_{\rm pred}$) 
values given in Table 2. 
}
\label{lower_limits}
\end{figure}

{\it PG\,0855+294}: The Villa-Elisa H\,{\sc i} spectrum shows a weak HVC at about --169 km\,s$^{-1}$. 
There is no corresponding Ca\,{\sc ii} detection. WM00 predicts  log($N_{\rm CaII}$)=11.57, compared with our 
limit of 11.16. The data imply that this HVC is further away than the stellar $z$-height 
of 4100 pc at a 1$\sigma$ level only. 
Fig. \ref{lower_limits} shows the result of a model fit using log$N$(Ca\,{\sc ii})=11.57, $b$=9.7 km\,s$^{-1}$ (estimated 
from the H\,{\sc i} profile), and $v_{\rm LSR}$=--168.7 km\,s$^{-1}$, 
superimposed on the observed spectrum. Part of the IV Arch at --27 km\,s$^{-1}$ is also detected in H\,{\sc i}. 
This has a corresponding Ca\,{\sc ii} K detection at --29 km\,s$^{-1}$, giving an upper $z$-height to this IVC (IV20 
from Kuntz \& Danly 1996) of 4100 pc.  

{\it EC\,09452--1403}: The Dwingeloo H\,{\sc i} spectrum shows a possible HVC at +120 km\,s$^{-1}$, 
but its brightness temperature is only 0.10 K. There is no corresponding Ca\,{\sc ii} detection. WM00 
predicts log($N_{\rm CaII}^{\rm WM00}$)=11.46, compared with our upper limit of 11.36. Hence the current 
data does not constrain the distance to this HVC, even assuming it is real and not baseline ripple. 
Fig. \ref{lower_limits} shows the result of a model fit using 
log$N$(Ca\,{\sc ii})=11.46, $b$=10.5 km\,s$^{-1}$ (estimated 
from the H\,{\sc i} profile), and $v_{\rm LSR}$=+120.0 km\,s$^{-1}$, superimposed on the observed spectrum. 

\begin{figure}
\includegraphics{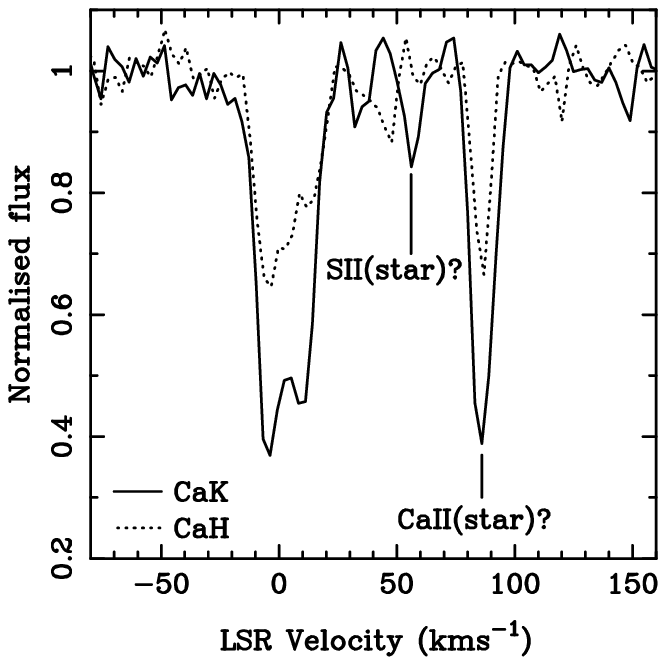}
\caption{Ca\,H and K spectra towards EC\,09470-1433.}
\label{ec09470spectrum}
\end{figure}

{\it EC\,09470--1433}: The Dwingeloo H\,{\sc i} spectrum shows a marginal IVC detection at 
                 +55 km\,s$^{-1}$ ($T_{B}^{\rm peak}$=0.25 K). The HVC detection 
                 at +114 km\,s$^{-1}$ is more secure. The optical absorption at +56 km\,s$^{-1}$ is 
                 likely to be stellar S\,{\sc ii} at $\lambda_{\rm air}$=3933.264\AA. If this feature were 
                 interstellar in nature, it would also be present in the Ca\,H spectrum shown in 
                 Fig. \ref{ec09470spectrum}. However, this is not the case. The detection at +86 km\,s$^{-1}$ 
                 is likely to be stellar Ca\,{\sc ii} K due to the velocity and velocity width being consistent 
                 with a stellar feature, plus the lack of obvious H\,{\sc i} at this position. 
                 Concerning the H\,{\sc i} feature at +114 km\,s$^{-1}$, there is no obvious associated Ca\,{\sc ii} K. 
                 WM00 predict log($N_{\rm CaII}^{\rm WM00}$)=11.35, compared with our upper limit of 10.98. 
                 Hence at the $\sim$ 1$\sigma$ level only, this HVC is likely to lie at a $z$-height exceeding 1400 pc. 
                 Fig. \ref{lower_limits} shows the result of a model fit using log$N$(Ca\,{\sc ii})=11.35, $b$=10.6 km\,s$^{-1}$ (estimated 
                 from the H\,{\sc i} profile), and $v_{\rm LSR}$=+114.0 km\,s$^{-1}$, superimposed on the observed spectrum. 

  {\it PG\,0955+291}: The Dwingeloo H\,{\sc i} spectrum shows an IV feature at 
  --29 km\,s$^{-1}$ extending to --70 km\,s$^{-1}$ with log($N_{\rm HI}$)=19.81. An 
  additional H\,{\sc i} detection at +115.5 km\,s$^{-1}$ is weak although appears 
  real. This small HVC has no corresponding Ca\,{\sc ii} absorption detection. Fig. \ref{lower_limits} 
  shows the result of a model fit to the Ca\,{\sc ii} spectrum using 
  log$N$(Ca\,{\sc ii})=11.45, $b$=7.5 km\,s$^{-1}$ (estimated from the H\,{\sc i} 
  profile), and $v_{\rm LSR}$=+115 km\,s$^{-1}$, superimposed on the observed spectrum.

  The Ca\,{\sc ii} K data (previously analysed by Ryans et al. 1997a) also show Ca\,{\sc ii} 
  absorption at --45 and --64 km\,s$^{-1}$ which were used to set an upper $z$-height limit of 4300 pc
  to the IVC IV18. 

{\it PG\,1008+689}: The Dwingeloo H\,{\sc i} spectrum shows H\,{\sc i} associated with the LLIV arch at 
              --44 km\,s$^{-1}$ with log($N_{\rm HI}$)=19.62$\pm$0.03. Ryans et al (1997a) also detected this 
              feature in their Lovell telescope data, with a velocity of --45.4 km\,s$^{-1}$ and log($N_{\rm HI}$)=19.55. 
              They used these data to place an upper distance limit to this IVC. In the H\,{\sc i} spectrum 
              there is only one component, whereas in the Ca\,{\sc ii} data there are components 
              at $\sim$--39 and --49 km\,s$^{-1}$. The total Ca\,{\sc ii} column density contained 
              in these two features is log($N_{\rm CaII}$)=11.86.

{\it EC\,10087--1411}: The Dwingeloo H\,{\sc i} spectrum shows a clear IVC at +63 km\,s$^{-1}$. 
                 The cloud is not seen in Ca\,{\sc ii} K absorption to a limit of 
                 log($N_{\rm CaII}$)=11.16 compared to the value of 11.66 predicted 
                 by WM00. Hence the current data put a lower $z$-height
                 limit of 620 pc to the current cloud, at a 1$\sigma$ limit only.

{\it EC\,11074--2912}: The intermediate-velocity feature at --48.6 km\,s$^{-1}$ merges in with LV gas. 
                       No Ca\,{\sc ii} absorption is seen below a velocity of --30 km\,s$^{-1}$.

{\it EC\,11507--2253}: The Dwingeloo H\,{\sc i} spectrum shows a possible IVC at +45 km\,s$^{-1}$ that 
                 merges into the LV gas. This feature is also weakly detected in optical absorption 
                 in the Ca\,{\sc ii} K and Ca\,{\sc ii} H spectra. If the two features probe the same
                 material, the IVC is closer than the (unknown) stellar distance. 

{\it PG\,1213+456}: The Dwingeloo H\,{\sc i} spectrum shows an IVC at --58 km\,s$^{-1}$ with 
                    log($N_{\rm HI}$)=19.55. The Lovell-telescope H\,{\sc i} data of Ryans et al. (1997a) 
                    shows a feature at --60 km\,s$^{-1}$ and log($N_{\rm HI}$)=19.40 which they 
                    used to determine the distance to IV17.              
 
{\it PG\,1243+275}: The Dwingeloo H\,{\sc i} spectrum shows an LV feature at --23 km\,s$^{-1}$ that 
               extends to --60 km\,s$^{-1}$. The Ca\,{\sc ii} K spectrum shows a corresponding 
               absorption feature at --40 km\,s$^{-1}$, blended in with lower-velocity gas. {\it If} these 
               are the same feature, then the IV gas must lie at a $z$-height of less than 6200 pc.

{\it LS 3510}: The Villa-Elisa H\,{\sc i} spectrum shows strong IV emission at +38 and +53 km\,s$^{-1}$. There are no 
               corresponding Ca\,{\sc ii} K detections which imply that the distance to these features is greater
               than that of the (unknown) stellar distance. 

{\it LS 3694}: The Villa-Elissa H\,{\sc i} spectrum shows strong IV emission at +41 and +55 km\,s$^{-1}$. There are no
               corresponding Ca\,{\sc ii} K detections which imply that the distance to these features is greater
               than that of the (unknown) stellar distance. 

{\it LS 3751}: The Villa-Elissa H\,{\sc i} spectrum shows strong IV emission at $\sim$ +58 km\,s$^{-1}$. There are no
               corresponding Ca\,{\sc ii} K detections which imply that the distance to these features is greater
               than that of the (unknown) stellar distance. A IVC is, however, detected in Ca\,{\sc ii} K absorption at
               +39 km\,s$^{-1}$, placing an upper limit to this feature at equal to the (unknown) stellar distance.              

{\it PG\,1725+252}: The Dwingeloo H\,{\sc i} spectrum shows an HVC at --154 km\,s$^{-1}$. There is no 
               corresponding feature seen on the low-resolution WHT spectrum. WM00 predict 
               log($N_{\rm CaII}^{\rm WM00}$)=11.71, compared with our upper limit of 
               11.28. Hence at the 1$\sigma$ level the HVC is at a $z$-height larger than 320 pc. 

{\it PG\,1738+505}: The Dwingeloo H\,{\sc i} spectrum shows an IVC at --98 km\,s$^{-1}$. There is no 
                    associated Ca\,{\sc ii} K absorption feature to a limit of $\sim$ 11.36, compared to 
                    the predicted value of 11.71. Hence these data say imply that the cloud is at a $z$-height
                    greater than the star of 510 pc at $<$ 1$\sigma$ level only. 

% This WHT spectrum does show weak 
% evidence of absorption at $\sim$ --60 km\,s$^{-1}$, although this is tentative. 
% If confirmed then this would put the IVC at a $z$-height less than 510 pc.

{\it HD\,341617}: The Dwingeloo H\,{\sc i} spectrum shows an HVC at --118 km\,s$^{-1}$, connected to LV
               material. No corresponding feature is seen in the Keck Ca\,{\sc ii} K spectrum.
               Fig. \ref{lower_limits} shows the result of a model fit using log$N$(Ca\,{\sc ii})=11.62, 
               $b$=14.9 km\,s$^{-1}$ (estimated from the H\,{\sc i} profile) and $v_{\rm LSR}$=--118 km\,s$^{-1}$, 
               superimposed on the observed spectrum.
               A {\it possible} IVC is detected in Ca\,{\sc ii} K at --58 km\,s$^{-1}$. This could
               be associated with the wing of gas extending from LV to HV material, but this is not 
               certain. If so, then this gas lies at a distance closer than the $z$-height of 2800 pc 
               estimated by Mooney et al. (2002).

\begin{figure}
\includegraphics{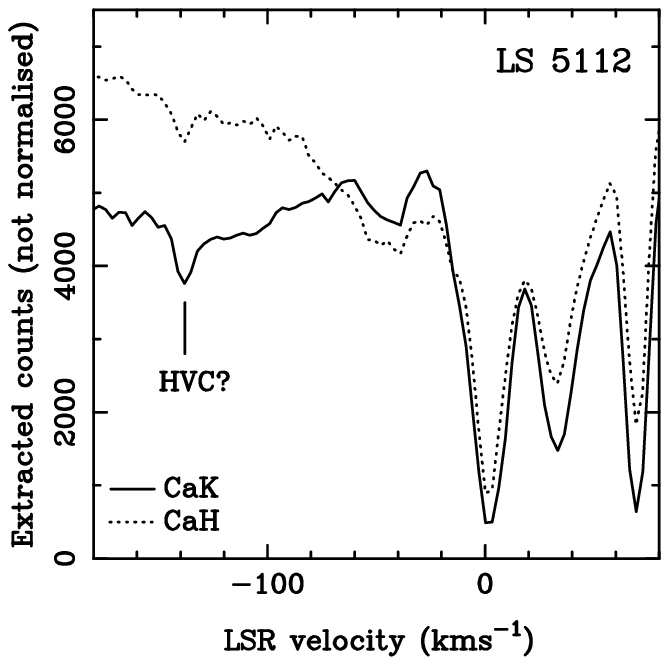}
\caption{Raw extracted Ca\,H and K spectra towards LS\,5112. The narrow-velocity feature at -138 km\,s$^{-1}$ 
marked ``HVC?'' is present in both spectra. }
\label{ls5112spectrum}
\end{figure}

{\it LS\,5112}: The Ca\,{\sc ii} K spectrum shows an absorption feature at $\sim$ --137 km\,s$^{-1}$ with FWHM 
                of 8.0 km\,s$^{-1}$. As the instrumental resolution is $\sim$7.0 km\,s$^{-1}$, the feature is 
                essentially unresolved. This narrow-width feature is superimposed on top of the Ca\,{\sc ii} K 
                stellar line at $v_{\rm LSR}$=--120 km\,s$^{-1}$, which has a FWHM of 58 km\,s$^{-1}$. 
                The narrow absorption feature is also present in the Ca\,H 
                spectrum as presented in Fig. \ref{ls5112spectrum}. If this is interstellar nature, it places an upper distance
                as equal to the (unknown) stellar distance. However, no H\,{\sc i} is detected at this velocity on the 
                Leiden-Dwingeloo survey. There is of course the possibility that the feature is circumstellar.  
                The positive-velocity features at high velocity are associated with normal 
                differential rotation. 

{\it EC\,19071--7643}: The Villa-Elisa survey shows an H\,{\sc i} wing to the LV gas that extends to 
                 $\sim$ --60 km\,s$^{-1}$. Due to the stellar velocity being $\sim$ --17 km\,s$^{-1}$ it 
                 is difficult to determine whether the wing also seen in Ca\,{\sc ii} K is stellar 
                 or interstellar in nature. Hence the current data are of no use in determining the 
                 distance to this IVC. 
                 
{\it EC\,19489--5641}: The Villa-Elisa survey H\,{\sc i} spectrum shows a IV feature at 
                 +36 km\,s$^{-1}$ that merges in with LV gas. There is no corresponding feature on 
                 the Ca\,{\sc ii} K spectrum and the stellar distance is unknown. 
                 There are narrow and weak features in the Ca\,{\sc ii} K 
                 spectrum at --128 km\,s$^{-1}$ ($\lambda_{\rm air}$=3931.91\AA), 
                 --88 km\,s$^{-1}$ ($\lambda_{\rm air}$=3932.41\AA) and --78 km\,s$^{-1}$ 
                 ($\lambda_{\rm air}$=3932.54). The --128 km\,s$^{-1}$ feature is likely to be stellar 
                 S\,{\sc ii} ($\lambda_{\rm air}$=3931.91\AA). For the other two features the nearest 
                 stellar lines could be S\,{\sc ii} at $\lambda_{\rm air}$=3932.30\AA, 
                 Ar\,{\sc ii} at 3932.55\AA \, or Sc\,{\sc i} at $\lambda_{\rm air}$=3932.60\AA, the latter 
                 of which would not be seen in a B-type spectrum (e.g. Hambly et al. 1997). There 
                 thus remains the possibility that these lines are interstellar although there is 
                 lack of associated H\,{\sc i}.

{\it EC\,19490--7708}: The Villa-Elisa survey shows a possible IVC at +43 km\,s$^{-1}$. The feature is not 
                 visible on the Ca\,{\sc ii} K spectrum. WM00 predict log($N_{\rm CaII}^{\rm WM00}$)=11.59, 
                 compared with our upper limit of 10.91. Hence at the 1.5$\sigma$ 
                 level, the current observations put a lower $z$-height limit of $\sim$ 800 pc towards 
                 this IVC. The feature at $\sim$ +175 km\,s$^{-1}$ is likely He\,{\sc i} at 
                 $\lambda_{\rm air}$=3935.9\AA.

{\it EC\,19596--5356}: The Villa-Elisa survey shows a possible IV feature at $\sim$ +38.7 km\,s$^{-1}$ that 
                 merges in with LV gas. The feature is also detected on the Ca\,{\sc ii} spectrum, although
                 it again is blended with LV gas. The log($N_{\rm CaII}$) value of 11.45 compares to 
                 that predicted by WM00 of 11.61. If the features are coincident, this IVC must be closer 
                 than that of the star. Assuming a He abundance of 10.96 with the estimated $T_{\rm eff}$=16500 K, 
                 log($g$)=4.0 leads to a (large!) stellar distance of 48.1 kpc or a $z$-height of 20 kpc (Lynn, unpublished 
                 results), hence the IVC must be closer than this. 

%
% Text removed in second iteration
%
% {\it EC\,20068--7324}: The Villa-Elisa survey has a weak LV feature at --23.8 km\,s$^{-1}$ extending to about 
%                 --50 km\,s$^{-1}$, although at a low level. This feature is not visible as Ca\,{\sc ii} K 
%                 absorption.  

{\it EC\,20089--5659}: The Villa-Elisa H\,{\sc i} spectrum shows strong IVC emission at +50 km\,s$^{-1}$.
                 There is no obvious Ca\,{\sc ii} K absorption at this velocity on-top of the broad stellar 
                 line. WM00 predict log($N_{\rm CaII}^{\rm WM00}$)=11.72 compared with our limit of $<$ 10.61. 
                 Hence it seems likely that this IVC is at a $z$-height exceeding $\sim$ 400 pc.

{\it EC\,20104--2944}: The optical spectrum shows Ca\,{\sc ii} absorption 
                 at +52 km\,s$^{-1}$ with FWHM $\sim$ 10 km\,s$^{-1}$ with no obvious associated H\,{\sc i}  
                 emission on the LDS. At this position, normal Galactic rotation predicts velocities of up to 
                 $\sim$ +25 km\,s$^{-1}$. The absorption is not stellar as EC\,20104--2944 has a 
                 projected rotational velocity of 50 km\,s$^{-1}$. The presence of optical absorption 
                 implies that the gas lies at a $z$-height of less than $\sim$ 1860 pc. Finally, the 
                 line at $\sim$ 140 km\,s$^{-1}$ is probably stellar He\,{\sc i} at 
                 $\lambda_{\rm air}$=3935.95\AA. 

{\it EC\,20411--2704}: The Dwingeloo survey shows low-level ($<$0.2 K) H\,{\sc i} from $\sim$ 
                 --30 km\,s$^{-1}$ to --150 km\,s$^{-1}$. This is not visible in Ca\,{\sc ii} K, 
                 although the expected absorption would in any case be too weak to measure. 
 
{\it M\,15 ZNG--1}: The Dwingeloo survey shows a previously-identified IVC at +70 km\,s$^{-1}$ whose distance 
                    is less than 3 kpc. At a resolution in H\,{\sc i} of $\sim$ 1 arcmin, Smoker et al. (2002) 
                    find that this IVC has a large log($N_{\rm CaII}$) value of 11.67 and log($N_{\rm HI}$) value 
                    towards this sightline of 18.70. The current paper adds 
                    nothing to the information concerning this IVC. 

{\it EC\,23169--2235}: The Dwingeloo survey shows tentative evidence for H\,{\sc i} from 
                 $\sim$ --30 km\,s$^{-1}$ to --100 km\,s$^{-1}$, although this could be baseline 
                 ripple. The feature in any case would be too weak to detect via Ca\,{\sc ii} 
                 absorption hence the current data say nothing about the distance towards this IVC. 

{\it PG\,2351+198}: The Dwingeloo survey shows a possible IVC at --55 km\,s$^{-1}$, although blended 
with LV gas. The feature in the Ca\,{\sc ii} K spectrum at --100 km\,s$^{-1}$ is likely He\,{\sc i} at 
$\lambda$=3935.95\AA. The feature at --55 km\,s$^{-1}$ on the optical spectrum is 
at a rest wavelength of $\sim$ 3936.55\AA. This could either be interstellar Ca\,{\sc ii} K 
in absorption, or much less likely, stellar Mn\,{\sc i} at $\lambda_{\rm air}$=3936.76\AA which is 
normally not seen in B-type stars.

\section{Distance limits for individual cloud complexes}
\label{distance}

In the following section we collate the measurements given in the previous section in 
order to attempt to use them to provide distance limits to known IVC and HVC complexes. 

\subsection{Complex C}

Complex C is a huge H\,{\sc i} feature visible in the Northern Hemisphere. Its distance is 
not well constrained; Wakker (2001) gives a firm lower $z$-height limit of 800 pc and a weak
lower $z$-height limit of 4300 pc. We note that previous observations towards the QSO PG\,1351+640 in 
Ca\,{\sc ii} have found log($N_{\rm CaII}$)=11.91 at log($N_{\rm HI}$)=18.86. Hence, our upper limits of 
log($N_{\rm CaII}$)$<$11.28 at log($N_{\rm HI}$)=19.47 for PG\,1725+252 ($v_{\rm LSR}$=--154 km\,s$^{-1}$), 
log($N_{\rm CaII}$)$<$10.62 at log($N_{\rm HI}$)=19.18 for HD\,341617 ($v_{\rm LSR}$=--118 km\,s$^{-1}$) and 
log($N_{\rm CaII}$)$<$11.16 at log($N_{\rm HI}$)=19.00 ($v_{\rm LSR}$=--169 km\,s$^{-1}$) for PG\,0855+294 
would appear to give weak lower $z$-height limits of 320, 2800 and 4100 pc, respectively 
towards the corresponding parts of this complex. 

\subsection{Complex K and an IVC towards it}

Complex K is a northern cloud which exhibits weak H$\alpha$ emission (Haffner et al. 2001) and has LSR 
velocities ranging from $\sim$--65 to $\sim$ --95 km\,s$^{-1}$ and an existing upper $z$-height limit of 
4500 pc. Its deviation velocity of $\sim$ --80 km\,s$^{-1}$ (Wakker 2001) puts it on the borderline 
between the normal IVC/HVC demarcation. Although a total of 8 of our stars intersect regions of this cloud, 
in only one of these sightlines is H\,{\sc i} seen at velocities associated with Complex K; at 
$v_{\rm LSR}$=--98 km\,s$^{-1}$ towards PG\,1738+505 at a $z$-height of 510 pc.
Towards this sightline, it is unclear as to whether or not absorption in Ca\,{\sc ii} K was detected (Fig. \ref{fig1}). 

In the same region of the sky, we detect marginal Ca\,{\sc ii} K absorption with $v_{\rm LSR}$=--45 km\,s$^{-1}$  
towards HD\,341617 at $z$=2800 pc. The absolute value of the velocity is probably too low for this IVC 
to be considered part of Complex K. In any case, H\,{\sc i} in emission is only marginally detected towards this 
sightline at this velocity. Concluding, the current data do not say anything definitive about the distance 
to Complex K or the possible IVC towards HD\,341617. 

\subsection{Complex gp}

Complex gp is a southern positive-velocity IVC with an LSR velocity of $\sim$ +70 km\,s$^{-1}$ 
and an existing distance bracket of 300--2000 pc. Parts of it are in the same area of 
the sky as the Magellanic Stream, although the velocities of the features are different. 
The cloud is a strong H$\alpha$ emitter and has tentatively been detected using {\sc iras} 
(Smoker et al. 2002). A total of 5 of our sample intersect regions of this cloud, however, 
only in the previously-known direction of M\,15 is H\,{\sc i} detected. Hence the current 
observations add nothing to the distance estimate of this cloud. 

\subsection{Complex GCN}

Complex CGN contains a number of small clouds in the region of the Galactic Centre, which 
have velocities in the range of $\sim$ --340 to --170 km\,s$^{-1}$ (Wakker \& van Woerden 1991 and 
refs. therein). One of our stars of unknown distance, LS\,5112, intersects the general vicinity 
of the cloud. However, although we see absorption in the spectrum at $\sim$ --138 km\,s$^{-1}$ 
(close to the stellar velocity of --120 km\,s$^{-1}$), that may be interstellar or circumstellar 
Ca\,{\sc ii} H and K, there is no associated H\,{\sc i} emission at this velocity. 

\subsection{Complex WA--WB}

A few of our sightlines are in the same part of the sky as complexes WA--WB (Wannier, Wrixon \& Wilson 1972). 
Towards EC\,09452--1403 (complex WB) and EC\,09470--1433 H\,{\sc i} HV gas is detected at $v_{\rm LSR}$=+120 and 
$v_{\rm LSR}$=+114 km\,s$^{-1}$, respectively, with log($N_{\rm HI}$)=18.65 and 18.97. Neither of these 
sightlines is detected in Ca\,{\sc ii} absorption. Due to its relative faintness, EC\,09452--1403 only 
imposes a weak lower $z$-height limit of 2470 pc, although the lack of Ca\.{\sc ii} to a 
limit of log($N_{\rm CaII}$)$<$10.98 towards EC\,09470--1433 means it is likely that this complex 
lies at a $z$-height exceeding 1400 pc. Finally, a possible IVC is seen in the H\,{\sc i} spectrum 
in the vicinity of EC\,10087--1411. This is not seen in absorption at the 1$\sigma$ level, hence 
if it exists, a weak lower $z$-height limit of 620 pc is present for this object.

\subsection{Complex WE and an IVC towards it}

Complex WE is a HVC at low-intermediate Galactic latitude  with velocities of $\sim$+110 km\,s$^{-1}$ 
compared to an expected differential galactic rotation of between 0 and --100 km\,s$^{-1}$ in this 
direction (Wakker 2001). The $z$-height was previously thought to be $<$ 3200 pc. Towards one of our sightlines, 
EC\,06387--8045, we find a weak H\,{\sc i} detection at +73 km\,s$^{-1}$, with a correspondingly tentative 
Ca\,{\sc ii} K absorption at +75 km\,s$^{-1}$. However, it is not clear that this cloud is 
associated with complex WE, which has a velocity on this part of the sky of $\sim$+120 km\,s$^{-1}$; it is 
likely that this object is just an unrelated IVC. If both H\,{\sc i} and Ca\,{\sc ii} features 
are real, and there is doubt about this, then the 
upper $z$-height for this cloud would be 1770 pc.  Note that additionally, in the H\,{\sc i} spectrum of 
EC\,06012--7810, there is a wing extending up to 150 km\,s$^{-1}$ that could be associated with Complex WE.
This is not visible in the Ca\,{\sc ii} 
spectrum, although given its broad nature it would be very difficult to detect. Other IVCs are also seen in 
H\,{\sc i} emission towards stars LS\,3510, LS\,3604, LS\,3694 and LS\,3751, with velocities of $\sim$ +40 to 
+60 km\,s$^{-1}$. None of these are seen in Ca\,{\sc ii} K absorption, indicating that these IVCs are further
than the associated (unknown) stellar distances. The only IVC seen in absorption is at a velocity of 
$\sim$ +39 km\,s$^{-1}$ towards LS\,3751. 

\subsection{Southern IVCs including the Pegasus-Pisces Arch}

Of the current sample, in only PG \,0009+036 and PG\,2351+198 do we see H\,{\sc i} at intermediate 
velocity towards the IV South map of Wakker (2001, Fig. 17). The former sightline lies towards the 
Pegasus-Pisces Arch (PPA; Wakker 2001). Our marginal Ca\,{\sc ii} 
detection at $v$=--43 km\,s$^{-1}$ towards this sightline would place this part of the IV gas at 
a $z$-height of $<$ 9000 pc. A higher SNR spectrum towards this star would be useful. 
Towards PG\,2351+198 at $z$=2470 pc Ca\,{\sc ii} K absorption is seen coincident with 
H\,{\sc i} emission at --55 km\,s$^{-1}$, thus providing an upper distance limit to this 
part of the Southern IVC complex, assuming the feature is not Mn\,{\sc i} in absorption. 

\subsection{The IV Arch and Spur}

The IV arch and Spur covers a large part of the Northern sky, and consists of H\,{\sc i} with velocities 
around +60 km\,s$^{-1}$ (Kuntz \& Danly 1996) and $z$-height of between 800 and 1500 pc. Many of 
our sightlines have previously been analysed by Ryans et al. (1997a,b). The only new observations 
that we have are detections of Ca\,{\sc ii} K towards PG\,1243+275 at --43 km\,s$^{-1}$, with an 
implied upper $z$-height limit of 6200 pc and detection of Ca\,{\sc ii} K towards PG\,0855+294 
towards IV20, giving an upper $z$-height limit of 4100 pc towards this cloud. These observations do 
not improve the previously-existing distance bracket towards the IV Arch or Spur. 

\subsection{The Magellanic Stream and IV clouds towards it}

The Magellanic Stream is an enormous H\,{\sc i} feature, spanning more than 100 degrees of 
both the northern and southern sky. It is thought to be tidal in nature, formed by the 
interaction of the Magellanic Clouds with the Milky Way, and velocities across it of 
$\approx$ --300 to +300 km\,s$^{-1}$ (Wakker 2001; Putman et al. 2003 and refs. therein).
Although a few of our sightlines intersect the Stream and are detected in the H\,{\sc i} spectra, 
no Ca\,{\sc ii} components are seen, due to the fact that the head of stream is at a distance 
of $\sim$ 55 kpc, compared with the stellar distances of $<$ 10 kpc. 

A few IVCs have been tentatively detected towards stream sightlines in the current dataset, 
although of course they need not be associated with it. EC\,19596--5356 shows H\,{\sc i} 
emission at +39 km\,s$^{-1}$ and possible Ca\,{\sc ii} absorption at +43 km\,s$^{-1}$, 
leading to a tentative upper $z$-height limit of $\sim$ 20 kpc. Similarly, EC\,20104--2944 shows 
Ca\,{\sc ii} absorption at 53 km\,s$^{-1}$ with no obvious associated H\,{\sc i}  
emission on the LDS. At this position, normal Galactic rotation predicts velocities of up to 
$\sim$ +25 km\,s$^{-1}$. 
  
\subsection{Previously uncatalogued HVCs}

In one of our sightlines, towards PG\,0955+291, there is a small, previously-uncatalogued 
HVC at $l,b$=200$^{\circ}$,+52$^{\circ}$ with $v_{\rm LSR}$=+115 km\,s$^{-1}$. Due to its
relative faintness in H\,{\sc i} ($\sim$4$\times$10$^{18}$ cm$^{-2}$), the current observations 
only put a weak lower $z$-height limit of 4300 pc towards this cloud 
(see Fig. \ref{lower_limits}).

%%%%%%%%%%%%%%%%%%%%%%%%%%%%%%%%%
\section{Summary and Conclusions}
\label{concl}
%%%%%%%%%%%%%%%%%%%%%%%%%%%%%%%%%

We have searched for interstellar absorption in the Ca\,{\sc ii} K line for traces of intermediate 
and high velocity clouds. A number of clouds were detected in this species, although, as normal 
in this kind of work, many showed no evidence for Ca\,{\sc ii} absorption in the spectra. Under assumptions 
concerning the abundance variation of the gas and changes in column density over the IHVC in 
question, the current data were used to estimate lower distance limits towards a number of IHVCs. 
Future work should follow up a number of the tentative distance estimates, by obtaining 
higher signal to noise spectra than is present in a number of the current stars and higher-resolution
H\,{\sc i} data. 

\section*{acknowledgements}
We would like to thank the staffs of Isaac Newton Group of telescopes, La Palma, Spain,
the Anglo-Australian Observatory, Coonababraran, Australia, the European Southern Observatory, 
Cerro Paranal, Chile (programme ID 67.D-0010A) and the W.M. Keck observatory, Hawaii, U.S.A., 
for help in taking these observations. HRMK, WRJR and RSIR thank {\sc pparc} for financial 
support for some of this work. BBL and CJM would like to thank the Department for Employment 
and Learning, Northern Irelend for funding. JVS would like to thank the APS division of Queen's University 
Belfast for hospitality as part of the visiting fellows programme, the European Southern 
Observatory for travel funds and D. E. Faria for useful comments. FPK is grateful to AWE 
Aldermaston for the award of a William Penney Fellowship. This research has made use of 
the {\sc simbad} database, operated at CDS, Strasbourg, France. Finally, we would like 
to thank the anonymous referee for many useful suggestions and corrections to the text. 

%%%%%%%%%%%%%%%
% References  %
%%%%%%%%%%%%%%%

{}

%%%%%%%%%%%%%%%%%
% Figure captions
%%%%%%%%%%%%%%%%%

\end{document}